\documentclass[useAMS,usenatbib,usegraphicx,normalem]{mn2e}
\usepackage{times}
\usepackage{xcolor} 

\newcommand{\lsim}{\lower0.6ex\vbox{\hbox{$ \buildrel{\textstyle <}\over{\sim}\ $}}}
\newcommand{\gsim}{\lower0.6ex\vbox{\hbox{$ \buildrel{\textstyle >}\over{\sim}\ $}}}
\newcommand{\beq}{\begin{equation}}
\newcommand{\eeq}{\end{equation}}

\DeclareRobustCommand{\kms}{\mathrm{km\,s^{-1}}}
\DeclareRobustCommand{\sbunits}{\mathrm{mag \,arcsec^{-2}}}

\title[Tidal features of Milky Way dwarf satellite galaxies]{
Tidal features of classical Milky Way satellites in a $\Lambda$CDM universe
}

\author[M.-Y. Wang et al.]
{M.-Y. Wang$^{1}$\thanks{Email:meiyu@physics.tamu.edu}, Azadeh Fattahi$^{2}$, Andrew P. Cooper$^{3}$, Till Sawala$^{4}$, Louis E. Strigari$^{1}$
\newauthor
, Carlos S. Frenk$^{3}$, Julio F. Navarro$^{2}$, Kyle Oman$^{2}$, and Matthieu Schaller$^{3}$\\
$^1$Department of Physics and Astronomy, Mitchell Institute for Fundamental Physics and Astronomy, Texas A\&M University, College Station, TX 77843-4242 \\
$^2$Department of Physics and Astronomy, University of Victoria, PO Box 1700 STN CSC, Victoria, BC, V8W 2Y2, Canada \\
$^3$Institute for Computational Cosmology, Department of Physics, University of Durham, South Road, Durham DH1 3LE, United Kingdom \\
$^4$Department of Physics, University of Helsinki, Gustaf Hallstromin katu 2a, FI-00014 Helsinki, Finland \\
}
\begin{document}

\date{\today}


\pagerange{\pageref{firstpage}--\pageref{lastpage}} \pubyear{2016}

\maketitle

\begin{abstract}

We use the APOSTLE cosmological hydrodynamic simulations to
examine the effects of tidal stripping on cold dark matter (CDM) subhaloes that
host three of the most luminous Milky Way (MW) dwarf satellite galaxies: Fornax, Sculptor, and Leo I. We identify simulated
satellites that match the observed spatial and kinematic distributions of stars in these galaxies, and track their evolution after infall. We find $\sim$ 30$\%$
of subhaloes hosting satellites with present-day stellar mass $10^6$-$10^8$ $M_{\odot}$
experience $>20\%$ stellar mass loss after infall. Fornax analogues have earlier infall times compared to Sculptor and Leo I analogues. Star formation in Fornax analogues continues for $\sim3$--$6$ Gyr after infall, whereas Sculptor and Leo I analogues stop forming stars $< 2$-$3$ Gyr after infall. Fornax analogues typically show more significant stellar mass loss and exhibit stellar tidal tails, whereas Sculptor and Leo I analogues, which are more deeply embedded in their host DM haloes at infall, do not show substantial mass loss due to tides. When additionally comparing the orbital motion of the host subaloes to the measured proper motion of Fornax we find the matching more difficult; host subhaloes tend to have pericentres smaller than that measured ones for Fornax itself. From the kinematic and orbital data, we estimate that Fornax has lost $10-20\%$ of its infall stellar mass. Our best estimate for the surface brightness of a stellar tidal stream associated with Fornax is $\Sigma \sim$ 32.6 mag $ {\rm arcsec^{-2}}$, which may be detectable with deep imaging surveys such as DES and LSST. 
\end{abstract}




\section{Introduction} 
\label{section:introduction} 

Milky Way (MW) dwarf spheroidal (dSph) satellite galaxies are among the most
dark matter (DM) dominated stellar systems in the Universe
\citep{Mateo_98,McConnachie2012,Walker:2012td}. High mass-to-light ratios
in these systems have been derived from their velocity dispersion under the
assumption that their stellar populations are in dynamical equilibrium.
However, this assumption of dynamical equilibrium has been called into question, and
must continue to be scrutinised as larger kinematic samples and deeper
photometric surveys become available. 

Around the MW, in addition to the clear case of Sagittarius
\citep{Ibata_etal94},  among the ``classical" dSphs only Carina
\citep{Munoz_etal06,Battaglia_etal12,McMonigal_etal14} and a few
``ultra-faint" dwarf galaxies (e.g.\ the Hercules dSph;
\citealt{Roderick_etal15}) show signs of having been affected by tides. Around M31 there is
evidence of ongoing tidal disturbance in several satellite
galaxies~\citep{Choi_etal02,Crnojevic_etal14}. Although tidal features are not
evident in the majority of dSphs, this may simply be because photometric
datasets are not sufficiently deep and wide enough to identify them. 
 
Models of galaxy formation in a $\Lambda{}$CDM cosmogony can only reproduce the present-day properties of the MW satellite system if a fraction of satellites experience significant mass loss as the result of tidal interactions with their host \citep{Li_etal10,Barber_etal15,Sawala_etal16}.
Tidal stirring may also be an effective mechanism for transforming disky field
dwarfs into objects with properties resembling dSphs \citep{Tomozeiu_etal16}.
Tidal stripping of satellite galaxies can also provide a unique test of the
nature of DM. For example, satellite galaxies are more vulnerable to tides if their gravitational potentials are shallow or cored \citep{Errani_etal15}, for example due to dark matter self-interactions \citep{Dooley_etal16}. Different dark matter scenarios, for example warm dark matter~\citep{Lovell_etal14} or late-decaying dark matter~\citep{Wang_etal14}, also
predict different subhalo assembly histories for a given dSph. ~\cite{Wang_etal16} show that the CDM subhaloes that best match the stellar mass and kinematics of Fornax are those that have been severely tidally stripped.
  
Several previous authors have used N-body simulations to examine the
observational signatures of tidal interactions on MW satellite
galaxies.~\cite{Munoz_etal08} examine whether the observed features of the
Carina dSph can be accounted for by a mass-follows-light model that includes
tidal disruption, finding that tidal stripping can strongly affect stars at
large radii.~\cite{Penarrubia_etal08} find that
once stars start to be stripped from a dSph, the dark-to-luminous mass ratio
increases, and consider the implications of this for the known
dSphs.~\cite{Battaglia_etal15} simulate subhaloes with orbits and structural
properties that match the Fornax dSph, and find that the structure and
kinematics of the stars are only mildly affected by tidal
interactions.~\cite{Klimentowski_etal07} study how unbound particles affect the
interpretation of stellar kinematics, with a specific application to Fornax.
~\cite{Read_etal06} also examine the effects of tidal stripping and tidal
shocking to the dSph velocity dispersion profiles.
  
In this paper, using high-resolution cosmological hydrodynamical simulations
from the APOSTLE project, we match simulated satellites, and hence subhaloes, to the observable properties of bright MW dSphs. We examine the
nature of faint stellar tidal features associated with these subhaloes and
make predictions applicable to real dSphs. Our study is the first attempt to examine
the formation of dSph tidal features in an cosmological context with full
hydrodynamical models. We
specifically focus on satellites in the stellar mass range of
$M_{*}=10^6$-$10^8 M_{\odot}$, which are very well
resolved by the APOSTLE simulations; the classical MW dSphs that fall within
this mass range are Fornax, Sculptor, and Leo I. We predict the surface
brightness of tidal streams associated with simulated dSphs, and compare to the
surface brightness of known diffuse stellar substructures in the halo of the
MW. We also consider the possibility that unbound stars contaminate
existing kinematic datasets for these galaxies. 

The outline of the paper is as follows. In \S~\ref{section:simulations}, we
briefly describe the simulations used in our analysis.
In~\S~\ref{section:methods} we review our procedure for constructing a subhalo
sample from the simulations by matching to observational data. In~\S~\ref{section:tidal effects} we show the general effects of gravitational tides on subhaloes, including the effects on dark matter and stellar mass loss, and changes to the galaxy stellar velocity dispersion. In~\S~\ref{section:dSph} and in~\S~\ref{section:tidal_debris} we present our results, including the overall
properties and mass assembly histories of the bright dSph analogues and the
detectability of tidal features around them. Finally we summarise our conclusions in
\S~\ref{section:conclusion}.

\section{Simulations}
\label{section:simulations}
We use sets of high-resolution LG simulations from the APOSTLE project (A Project Of Simulating The Local Environment). For details we refer the
readers to the original simulation papers \citep{Fattahi_etal16, Sawala_etal16}.
Here we briefly describe some basic properties of the simulations. 

The simulations were performed with the code developed for the EAGLE (Evolution
and Assembly of GaLaxies and their Environments) project
\citep{Schaye_etal15,Crain_etal15}. The Eagle code is a modified version of
P-Gadget-3, which is an improved version of the publicly available Gadget-2
code (Springel 2005). For our study, we utilize both the medium-resolution (MR) and high-resolution (HR) runs of the
APOSTLE simulations (corresponding to L2 and L1 resolution in \citet{Sawala_etal16}). The HR runs consist of three Local Group realizations (AP-1, AP-4, and AP-11
 in \cite{Fattahi_etal16}), which contain six MW-mass host haloes with
$M_{\rm 200} = 1.66\times 10^{12} $ $M_\odot$, $1.10\times 10^{12} $ $M_\odot$,  $1.38\times
10^{12}$ $M_\odot$, $1.35\times 10^{12} M_{\odot}$, $0.99\times 10^{12} M_{\odot}$, and $0.80\times 10^{12} M_{\odot}$ at $z=0$. $M_{\rm 200}$ is defined as the halo mass enclosed within a density contour 200 times greater than the
critical density of the Universe. The primordial gas particle mass of AP-1 (AP-4, AP-11) is $0.99$ $(0.49, 0.84)$ $\times  10^4 M_{\odot}$; the dark matter particle
mass is $5.92$ $(2.92, 5.02)$ $\rm \times 10^4 M_{\odot}$.  The maximum gravitational
softening length is 134~pc for all three HR runs. 

We also include the 12 Local Group pairs of MR runs in our analysis. The total mass of each pair (i.e., the sum of the virial masses
of the two primary haloes) is in the range $\log_{10} \,M_{\mathrm{tot}}/\mathrm{M_\odot} = [12.2, 12.6]$. The primordial gas particle mass spans the range of $1.09-1.27$ $\rm \times 10^5  M_{\odot}$, and the dark matter particle
mass range is $6.52$--$7.59 \rm \times 10^5  M_{\odot}$. The maximum gravitational softening length is 307~pc for all MR runs. The APOSTLE simulations were performed in the WMAP-7 cosmology \citep{Komatsu_etal11}, with density parameters at $z
= 0$ for matter, baryons and dark energy of $\Omega_{\rm m}= 0.272$,
$\Omega_{\rm  b} = 0.0455$ and $\Omega_{\Lambda}=0.728$, respectively, Hubble
parameter $\rm H_0 = 70.4 \mathrm{km \, s^{-1}} \rm Mpc^{-1}$, linear power
spectrum amplitude on the scale of $8h^{-1} \mathrm{Mpc}$ $\sigma_8$ = 0.81 and
power-law spectral index $n_s=0.967$.  Dark matter haloes in both simulations
were identified using the friends-of-friends (FoF) algorithm.
Each FoF halo was iteratively decomposed into self-bound subhaloes using the
SUBFIND algorithm \citep{Springel_etal01}.

\section{Methodology}
\label{section:methods}

In this section we outline how
we identify subhaloes in APOSTLE simulations that provide good matches to the observed
stellar kinematics of particular dSphs. We review the procedure outlined
in~\cite{Wang_etal16}, based on the Jeans formalism, which we follow closely
here, and also the dynamical mass estimation in~\cite{Fattahi_etal16b} that we adopt for selecting MR analogues. We then describe how we identify tidally stripped stars previously associated with simulated dSphs. 

\subsection{Selecting simulated analogues of real dSphs}
\label{subsection:selection}

We focus on MW satellites in the stellar mass range $\rm 10^6 M_{\odot}\le
M_{\star} \le 10^8 M_{\odot}$. Three of the so-called classical MW dSph
satellite galaxies have a stellar mass in this range (assuming a stellar mass-to-light ratio of order unity): Fornax (absolute magnitude $M_{V}=-13.2$), Sculptor ($M_{V} =-11.1$) and Leo I ($M_{V}=-11.9$) \citep{Mateo_98}. These three dSphs have similar line-of-sight (LOS) stellar
velocity dispersion $\langle\sigma_* \rangle \sim$ $9$--$11~\kms$
\citep{Mateo_etal08,Walker_etal09c}.

We consider a sequence of increasingly stringent conditions to select potential
bright dSph analogues in the simulations, as follows:

\begin{enumerate}

\item

Select subhaloes with stellar masses broadly consistent with the observed dSph
luminosity, assuming a stellar mass-to-light ratio $M/L= 0.5$--$2.0$. We choose
the acceptable Fornax stellar mass range to be $0.6$--$6.0\rm \times10^7
M_{\odot}$. For both Sculptor and Leo I the corresponding range is
$1.0$--$6.0\rm \times10^6 M_{\odot}$. Observational estimates of stellar mass
are $\sim1.7$--$4.3\times10^7 M_{\odot}$ for Fornax, $\sim0.2\times10^7
M_{\odot}$ for Sculptor and $\sim0.5\times10^7 M_{\odot}$ for Leo~I
\citep{Wolf_etal10,McConnachie2012,deBoer_etal12}.

\item 

Apply the Jeans formalism to select HR subhaloes with gravitational potentials that
match the observed stellar kinematics, using the methods described
in~\S\ref{subsection:jeans}. The spatial resolution of the MR simulations is not high enough to resolve the subhalo potential within the dSph half-light radius, so instead we use ${V_{\rm 1/2}}$ (the circular velocity at the half-light radius), as described in \citet{Fattahi_etal16b}, to constrain the stellar kinematic properties of our dSphs in these simulations.

\item 

Determine how well the orbits of these subhaloes match those derived from the
observed mean radial velocity and proper motion of the real dSphs by selecting those analogues at least with comparable orbital pericentres.

\end{enumerate} 

\begin{figure}
\includegraphics[height=8.9cm]{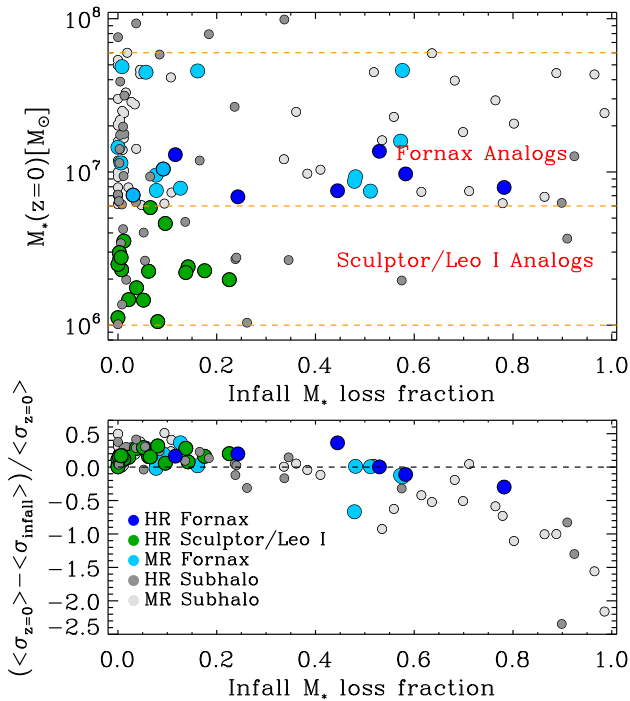}
\vspace{1em} 
\caption{ 
{\it Upper panel}: Fraction of stellar mass lost since infall for all luminous subhaloes with $10^6 \le M_* \le 10^8 M_{\odot}$. The \textit{y} axis gives the stellar mass at $z=0$, which may not equal the infall stellar mass due to mass loss and star formation after infall. {\it Lower panel} : Fractional difference between the stellar velocity dispersion at infall and at present time as a function of stellar mass loss. The dark blue points represent HR Fornax analogues, and the light blue points MR Fornax analogues. The green points represent HR Sculptor/Leo analogues. Dark (light) grey points show the rest of the subhaloes from HR (MR) simulations. 
}
\label{fig:ms_loss}
\end{figure}

\subsection{Fitting dSph stellar kinematics}
\label{subsection:jeans}

To identify subhaloes in the HR simulations with observed stellar kinematic and
photometric data to match Fornax, Sculptor, and Leo I, we use the spherical Jeans
equations, allowing for a constant but non-zero anisotropy in the stellar velocity
dispersion. We follow a similar approach to that outlined
in~\cite{Wang_etal16}, to which we refer the reader for more details. 

We begin by identifying subhaloes in the simulations with the halo finder
SUBFIND. For each subhalo, the self-bound dark matter distribution
is determined as a function of radius. Given the potential of each subhalo, the
parametrised stellar distribution, and velocity anisotropy, we solve the
spherical Jeans equations to determine the line-of-sight velocity dispersion
assuming that the potential is spherically symmetric, dispersion-supported, and
in dynamical equilibrium.  The derived velocity dispersion is then fitted to
the stellar kinematic data by marginalizing over model parameters via a Markov
Chain Monte Carlo (MCMC) method to determine the best fit values. 

The dSph stellar kinematic data we use are line-of-sight velocity
measurements from the samples described in~\cite{Walker_etal09b}. In this sample
we consider only stars with $>$ 90$\%$ probability of membership. We bin the
velocity data in circular annuli and derive the mean line-of-sight velocity in
each annulus as function of projected radius. The line-of-sight velocity
dispersion and associated error in each annulus are calculated following the method
described in~\cite{Strigari_etal10}.

Since MR subhaloes have poor resolution of the potential within the half-light radius of the dSphs we study, we adopt the approach described in \cite{Fattahi_etal16b} to 
identity the kinematic properties of these subhaloes. We select MR subhaloes that match the observed value of ${ V_{\rm 1/2}} = \sqrt{G{\rm M_{1/2}}/{\rm r_{1/2}}}$, which is the circular velocity at the half-light radius ${\rm r_{1/2}}$ with estimated enclosed mass ${\rm M_{1/2}}$. The values obtained for each Milky Way classical dSph are listed in \cite{Fattahi_etal16b}. The radii where V$_{\rm 1/2}$ is measured in the MR runs are smaller than the convergence radius defined by \citet{Power_etal03}. To correct for this numerical resolution, we use the approach described in Appendix A of \citet{Fattahi_etal16b}.  For example, to find Fornax analogues we select central\footnote{Central refers to the most massive structure of a given FoF group} galaxies in the stellar mass range of interest and measure the level of correction for their $V_{1/2}$. We then apply an average correction to $V_{1/2}$ of satellite galaxies in the same stellar mass range.  

\begin{figure*}
\includegraphics[height=7.8cm]{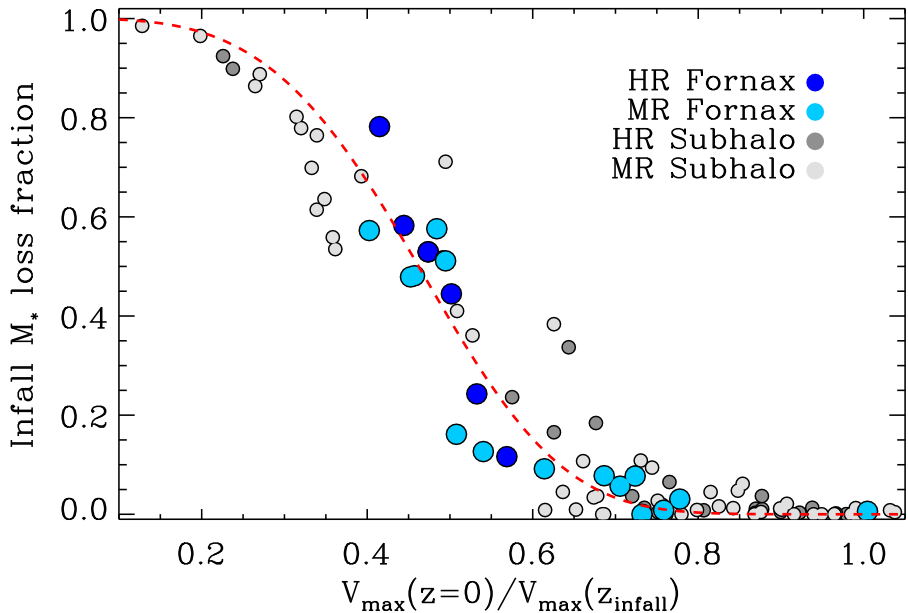}
\vspace{1em} 
\caption{Fraction of stellar mass lost since infall as a function of ${V_{\rm max}}(z=0)/{ V_{\rm max}}(z_{\rm infall})$ for subhaloes with stellar mass $6\times10^6 \le M_* \le 10^8 M_{\odot}$. The dark blue points represent the HR Fornax analogues, and light blue points MR Fornax analogues. The dark (light) grey points show the rest of the subhaloes from the HR (MR) simulations in this stellar mass range. The red dashed line illustrates the result of our fit of Eq.~\ref{eq:ms_loss} to these data.
}
\label{fig:dm_ms_loss}
\end{figure*}

\subsection{Tracing stellar tidal features}
\label{subsection:tidal stream}

An important part of our analysis is to determine the mass of dark matter and stars lost by a given dwarf satellite galaxy after it falls into its
Galactic-analogue host halo and the corresponding evolution of stellar tidal
features. To do this, we identify each dwarf galaxy at its time of infall
(using a merger tree algorithm that links haloes and subhaloes between
simulation outputs). We then consider the set of all stellar, gaseous and dark
matter particles that were gravitationally bound to this galaxy at any time
after the infall time. We trace all these particles down to $z=0$. If any new
particles are added to the system by ongoing star formation or the
gravitational capture of smaller systems between the infall time and the
present day, these will not be included in our procedure. It is then trivial to calculate, for example, the fraction of stellar mass lost since infall, and to separate particles at any time into those that have been stripped and those that remain bound to the satellite.

\section{Effects of Tides on Subhaloes}
\label{section:tidal effects}
In the upper panel of Fig.~\ref{fig:ms_loss} we plot the fraction of stellar
mass that has been lost since infall for all luminous  ($\rm 10^6 \le M_{\star}
\le 10^8 M_{\odot}$) subhaloes. This fraction is computed using only
those star particles that are bound to the subhalo at the time of infall and therefore does not include any stars formed (or accreted) after infall.  This additional stellar mass is, however, included in the present-day stellar mass plotted on the $y$-axis of Fig.~\ref{fig:ms_loss}. We use different colours to highlight HR and MR analogues of MW dSphs that we identify on the basis of their stellar kinematics (dark blue for HR Fornax, light blue for MR Fornax, and green for HR Sculptor/Leo I analogues). The rest of the subhaloes are shown as grey circles (dark grey for HR, and light grey for MR). We will describe the properties of these dSph analogues in detail in ~\S\ref{section:dSph}.

In general we find that present-day subhaloes with stellar mass in the range we study have experienced very different degrees of stellar mass loss. This reflects the diversity of their orbital properties and dark matter halo masses and sizes.
We find that 36 out of 127 ($\sim 28 \%$) subhaloes (taking HR and MR together) in the present-day stellar mass range shown have lost $\ge20\%$ of the stellar mass they had at infall. Among the subhaloes we identify as bright dSph analogues (of both Fornax and Sculptor/Leo), 11 out of 39 ($\sim 28 \%$) have lost more than $20 \%$, and majority of them are Fornax analogues. Some of these subhaloes have undergone even more significant stellar stripping, with mass loss $\ge40\%$ for $\sim 19 \%$ of all subhaloes and $\sim 23 \%$ of dSph analogues.  Notably, 10 out of 21 Fornax analogues (both HR and MR) have stellar mass loss fractions $\ge 20\%$. On the other hand, all Sculptor/Leo I analogues have mild stellar mass loss, $\lsim  20\%$. This result agrees with other studies of tidal effects on Milky Way dSphs using the APOSTLE simulations \citep{Sawala_etal16,Fattahi_etal16b}. These studies show that present-day haloes that have ${V_{\rm max}} < 30 \, \mathrm{km\,s^{-1}}$ and host observable dwarf galaxies have been affected by tidal stripping even more strongly than typical satellite haloes of the same mass. Similar findings are also shown in \cite{Sawala_etal16}, in particular dwarf galaxies in APOSTLE simulations are biased towards stronger stripping compared to haloes that did not form stars. As pointed out in \cite{Sawala_etal16}, this is an important factor in explaining why these simulations do not demonstrate the "too big to fail" problem posed by \citet{Boylan-Kolchin_etal11}. \cite{Fattahi_etal16b} also show that tidal stripping may have significantly reduced the dark matter content of several Milky Way dSphs. Fornax is a clear case for which the value of ${\rm V_{1/2}}$ is systematically lower than that expected for satellites with comparable luminosity (also see \citet{Cooper_etal09,Guo_etal15} for similar arguments).

\begin{figure*}
\includegraphics[height=7.4cm]{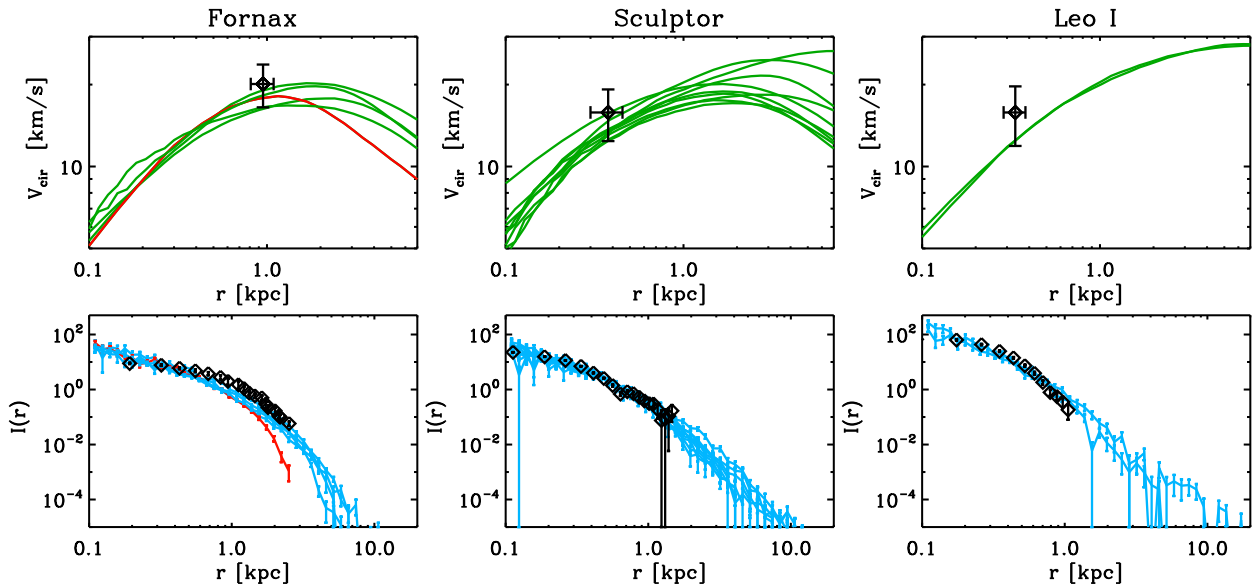}
\caption{
Circular velocity curves (upper panels) and stellar surface density profiles
(lower panels) for those subhaloes that match the Fornax (left), Sculptor
(middle), and Leo I (right) data at $z=0$. The derived ${\rm V_{ 1/2}}$ with uncertainties of $10-90^{th}$ percentile intervals for each dSph from \citet{Fattahi_etal16b} is marked by a black diamond in the upper panels. The photometric data
(black diamonds) in the lower panels are from \citet{Coleman_etal05} (Fornax,
lower left), \citet{Battaglia_etal08} (Sculptor, lower middle), and
\citet{Smolcic_etal07} (Leo I, lower right). The normalization of the projected
stellar density profiles is arbitrary, and the profiles from simulations are
projected along an arbitrary direction. The red lines in the Fornax panels
highlight the subhalo, which suffers the greatest amount of stellar mass
loss ($\sim$ 78 $\%$) among the analogues we identify (see
Fig.~\ref{fig:ms_loss}).
}
\label{fig:vcir}
\end{figure*}

In the lower panel of Fig.~\ref{fig:ms_loss} we show the fractional change in \textit{stellar} velocity dispersion between the infall time and the present day, as a function of the fraction of total mass lost since infall. We can interpret these results in the context of previous studies which showed how tidal forces and mass loss affect the stellar velocity dispersion of dwarf satellite galaxies \citep[e.g.][]{Penarrubia_etal08}. For modest amounts of total mass loss ($\le 20\%$) we find an increase in stellar velocity dispersion by up to $50\%$.  This is likely due to a combination of tidal heating, ongoing star formation and (in a small number of cases) ongoing accretion of stars from bound companions. For subhaloes that lose $\ge 90 \%$ of their total mass after infall, the stellar velocity dispersion drops drastically. In intermediate cases
(total mass loss fractions of $20$--$70\%$) the change in stellar velocity dispersion follows a smooth evolution between those two regimes with some scatters, reflecting
the changing balance between the different effects.

It is expected that the effects of tides on stellar components are highly correlated with the degree to which the dark matter halo is tidally stripped. The galaxies are embedded deeply inside the dark matter potential wells, thus subhaloes will typically experience significant dark matter halo mass loss before the main stellar distribution begins to show signs of tidal disturbance. In Fig.~\ref{fig:dm_ms_loss} we show the infall stellar mass loss fraction as a function of the ratio of subhalo circular velocity at the present day to that at infall, ${V_{\rm max}}(z=0)/{V_{\rm max}}(z_{\rm infall})$, for subhaloes (both HR and MR) with stellar mass $6\times10^6 \le M_* \le 1\times10^8 M_{\odot}$. We verify that the tidal effects on the dark matter and stellar components of subhaloes are consistent with this picture: there is no significant stellar tidal stripping (e.g. $\gsim 10 \%$) until the dark matter halo is severely stripped (e.g. ${V_{\rm max}}(z=0)/{V_{\rm max}}(z_{\rm infall}) \lsim 70 \%$). We find that the relation between infall stellar mass loss fraction ($f_{\rm *, loss}$) and change in ${V_{\rm max}}$ is well represented by the following fitting function:
\beq
f_{\rm *, loss}={\rm exp}(-\gamma x^{\alpha}),
\label{eq:ms_loss}
\eeq
where $x={V_{\rm max}}(z=0)/{V_{\rm max}}(z_{\rm infall})$. The best fit values of the parameters are $\gamma = 13.32$ and $\alpha = 3.83$. The fit is shown by the red dashed line in Fig.~\ref{fig:dm_ms_loss}.

\section{Properties of dwarf spheroidal analogues}
\label{section:dSph}

In this section we discuss the properties of the simulated bright dSph analogues that are either indicative of or determined their tidal stripping, including their present-day densities and velocity profiles, the evolution of their stellar, gaseous and dark mass before and after infall, and their orbital motions, which we compare to those estimated for their observed counterparts.


\subsection{Density and velocity profiles}
\label{subsection:properties}

In upper panels of Fig.~\ref{fig:vcir} we show the circular velocity profiles
of $z=0$ subhaloes in the HR runs that match the observed surface density of giant stars in Fornax, Sculptor, and Leo I and corresponding radial velocity measurements. Here, only these internal properties are used as constraints . The estimates of circular velocity at the half-light radius ($\rm V_{1/2}$) for each dSph (from \citealt{Fattahi_etal16b}, marked by black diamonds with uncertainties of $10-90^{th}$ percentile intervals) agree well with the results of our full Jeans analysis. The projected stellar
density profiles also agree well with observations (shown in the lower panels of Fig.~\ref{fig:vcir}, as black diamonds with Poisson errors). The observed
star count (number density) profiles are taken from \citet{Coleman_etal05}
(Fornax, lower left panel), \citet{Battaglia_etal08} (Sculptor, lower middle
panel), and \citet{Smolcic_etal07} (Leo I, lower right panel). We have scaled the stellar mass density profiles from the simulations to match the amplitude of the observed number density profiles of bright giants. For the simulated Fornax circular velocity curves, the maximum circular velocity
${V_{\rm max}}$ at $z=0$ is in the range $17$--$20\,\mathrm{km\,s^{-1}}$. In contrast to Fornax, the
Sculptor and Leo I analogues have a wider ${ V_\mathrm{max}}$ range with values up to $28 \, \kms$.
These ${V_\mathrm{max}}$ ranges agree well with previous works 
\citep[e.g. ][]{Penarrubia_etal08,Strigari_etal10,Boylan-Kolchin_etal12,Sawala_etal16,Fattahi_etal16b}
even though some of those analyses were carried out using dark matter-only simulations or assuming NFW potentials. The Fornax stellar density profiles
steepen beyond $r>1$ kpc, whereas Sculptor and Leo I have somewhat shallower
profiles.  We find that the average
stellar velocity dispersion of the bound star particles is $\sim 9$--$11\, \kms$, which agrees well with the observed values. This
indicates that those systems are in equilibrium and hence confirms
that our use of the Jeans formalism is appropriate for the simulated systems. We note that Sculptor and Leo I have very similar stellar kinematic properties and light profile shapes. The sets of Sculptor and Leo I analogues have many subhaloes in common. Therefore, we do not distinguish between them in the rest of the paper, except in the discussion of their orbital properties.

\begin{figure*}
\includegraphics[height=8.3 cm]{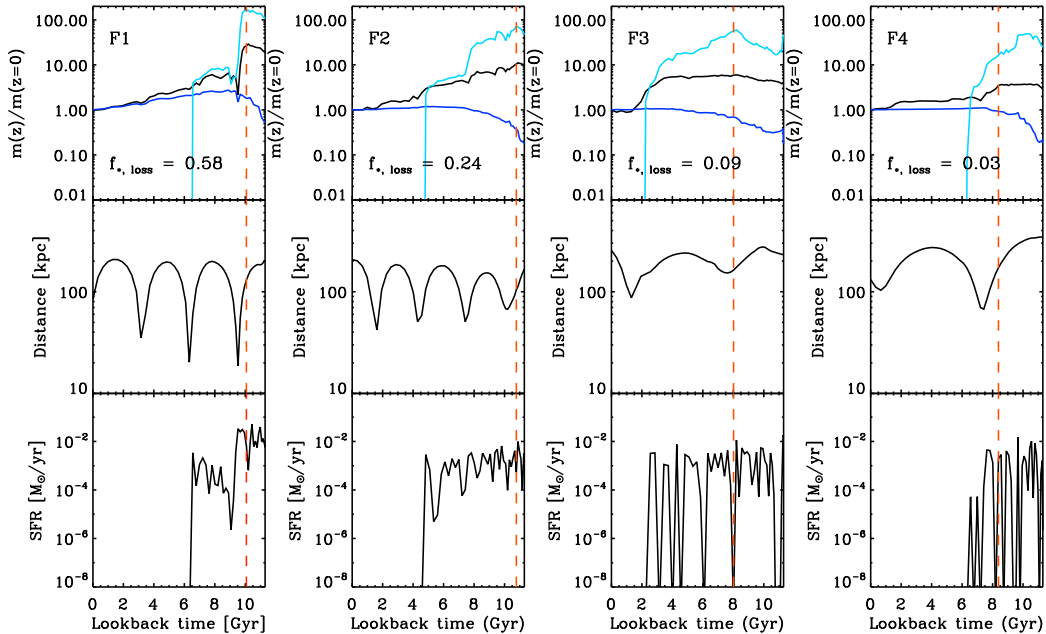}
\caption{ 
Mass assembly histories (upper row), distance to the host MW halo centre (middle row), and star formation rate for four Fornax analogues as a function of time. Each column shows the properties of one analogue (F1-F4). In the upper row, the black solid lines mark the redshift evolution of the subhalo total mass, dark blue lines are for the stellar component, and cyan lines are for the gas mass. The gas mass is normalized to the stellar mass at $z=0$. The red dashed lines indicate the subhalo infall time to their host haloes. The infall stellar mass loss fractions ($f_{*, loss}$) are labeled for each analogue. The analogues shown here are chosen to have different value of $f_{*, loss}$ (high to low from left to right).
}
\label{fig:fornax_mz}
\end{figure*}

\begin{figure*}
\includegraphics[height=8.5 cm]{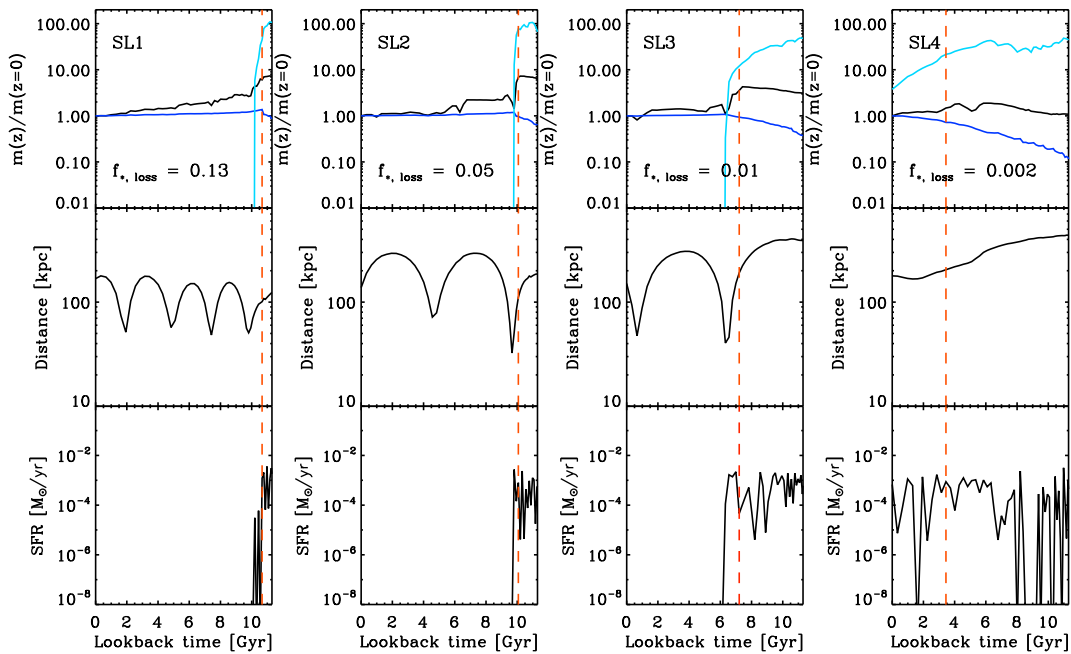}
\caption{ 
The same as Fig.~\ref{fig:fornax_mz}, except here for four Sculptor and Leo I analogues (SL1-SL4).
}
\label{fig:scu_mz}
\end{figure*}

In the leftmost panels of Fig.~\ref{fig:vcir} we highlight in red the
circular velocity and density profile of one particular Fornax analogue.
The density profile of this analogue is significantly shallower than those of
the other analogues and is a relatively poor fit to the observed Fornax light
profile. This analogue also has a smaller value of ${r_\mathrm{max}}$ (the radius corresponding
to ${V_\mathrm{max}}$) than the other Fornax analogues. These features
indicate that this subhalo has experienced a very significant amount of tidal stripping; we show in the next section that it has lost $\sim 78\%$ of its stellar mass at infall. As noted by \cite{Penarrubia_etal08}, even though stars deeply embedded within DM haloes are quite resilient to disruption, the loss of a significant fraction of their total mass nevertheless leads to an increase in the concentration of their light profiles. This seems to be the
case for our analogue.

\subsection{Mass evolution}
\label{subsection:infall}

 In the upper panels of Fig.~\ref{fig:fornax_mz} we show the evolution of total mass (black lines), stellar mass (dark blue lines) and gas mass (light blue lines) for four Fornax analogues, measured relative to their values at $z=0$ (The gas mass is normalized to the stellar mass at z = 0). These are computed by summing the masses of all particles gravitationally bound to the particular halo at each instant according to the SUBFIND algorithm\footnote{The assignment of particles to subhaloes in this algorithm is sensitive to the density contrast and velocity of the subhaloes relative to their host halo.
 Some material (in extreme cases the whole subhalo) may be considered unbound near pericentre but later considered bound again as the subhalo moves back
 towards apocentre.  This gives rise to transient `dips' in the bound mass near apocentres, which are visible in Fig. 4 and Fig. 5.  This feature of the SUBFIND algorithm does not affect our discussion here. In most cases the extent of the subhalo according to the definition used by SUBFIND is a good approximation to the true subhalo tidal radius.}. 

\begin{figure*}
\includegraphics[height=5.75cm]{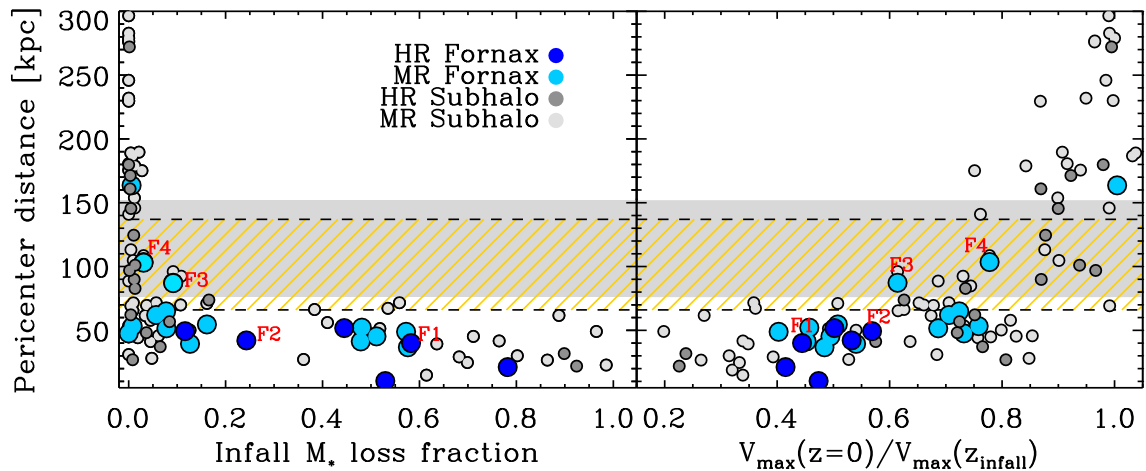}
\vspace{1em}
\caption{ 
Pericentre distance as a function of stellar mass loss fraction (left panel) and ${V_{\rm max}}(z=0)/{V_{\rm max}}(z_{\rm infall})$ (right panel) for subhaloes with stellar mass $6\times10^6 \le M_* \le 6\times10^7 M_{\odot}$, which is the luminosity range of our Fornax analogues. The colour scheme of the points is the same as Fig.~\ref{fig:dm_ms_loss}. The yellow shaded areas mark the 95$\%$ confidence interval range of orbital perigalacticon from \citet{Piatek_etal07}, and the grey shaded areas are derived perigalacticon 95$\%$ confidence interval range with Galactic potentials derived from APOSTLE simulations. The same observed Fornax velocity information (see \citet{Piatek_etal07}) is adopted in all perigalacticon calculations here. (see \S~\ref{subsection:orbit} for details).
}
\label{fig:dm_ms_peri}
\end{figure*} 

We find that the subhaloes of many Fornax analogues show large differences,
typically an order of magnitude, between their mass at $z=0$ and the maximum
total mass they reach over their lifetime. That peak mass occurs around the time of infall, which for these objects occurs at a lookback time of
$\sim8$--$10$ Gyr. Those analogues also show stellar mass loss fraction $\gsim 10\%$. The lower panels of Fig.~\ref{fig:fornax_mz} show the evolution of the star formation rate (SFR). These systems fluctuate around a low average SFR at high redshift. In most cases their SFR is not notably affected by infall or even the first pericentric passage, such that the peak stellar mass is reached after infall. Instead, star formation ends suddenly after 2--3 pericentres (3--6 Gyr) in response to an extremely rapid and near-total loss of gas from the system at that time (seen in the cyan lines in the upper panels). This behaviour is common to the majority of the Fornax analogues. 

The evolution of Fornax analogues after infall contrasts strongly with that of
Sculptor and Leo I analogues, four examples of which are shown in
Fig.~\ref{fig:scu_mz}. These systems have less dramatic differences between
peak mass and current mass and their infall times are typically more recent
($\sim 4$--$9$ Gyr ago), consistent with previous studies of the likely infall times of different MW dSphs~\citep{Rocha_etal12}. Their gas content and SFR fall to zero more quickly after infall, around the first apocentre unless their orbit stays far from the galactic centre the whole time (e.g. SL4).

The differences seen in the post-infall evolution of gas mass and the effects on the star formation histories of different analogues are interesting, and reminiscent of well-established differences in the stellar populations of observed dSphs (notably the presence of intermediate-age stars in Fornax). The catastrophic removal of gas is likely to be the result of ram-pressure stripping \citep{Gunn_etal1972}. 

Comparing different panels in Fig.~\ref{fig:fornax_mz} we also find that the level of tidal stripping is correlated with orbital properties, such as pericentre distance, number of  pericentre passages, and infall time. 

\subsection{Orbital Motions}
\label{subsection:orbit}

We now compare the orbital motions of our analogue subhaloes with the observed
motions of real dSphs, and thus determine whether the tidal forces on the
simulated galaxies are representative of those experienced by real dSphs.

\begin{figure*}
\includegraphics[height=9.5 cm]{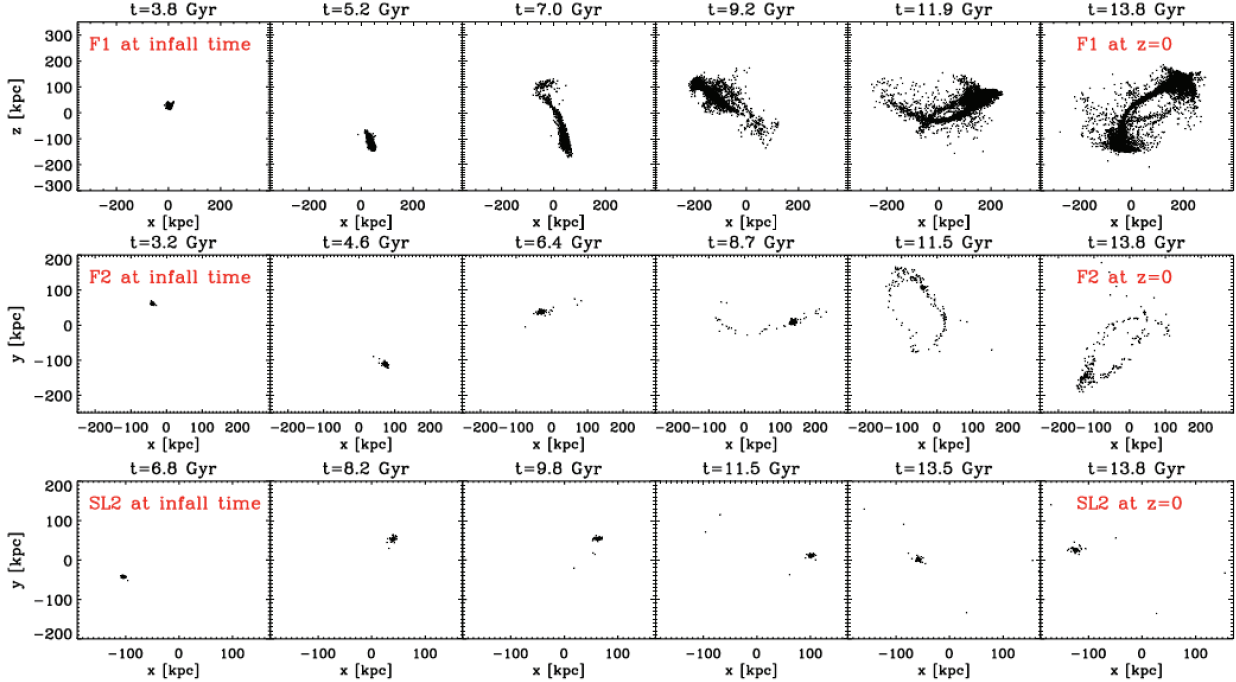}
\caption{ 
Projected spatial distribution of the star particles from two Fornax analogues
(upper two rows, F1 and F2) and one Sculptor/Leo I analogue (lower row, SL2) as a
function of the age of the Universe. These images include all star particles that are
gravitationally bound to those dSphs at their infall time or at any time
thereafter. For both Fornax analogues, tidally stripped star particles can be
seen to extend far beyond the main bodies of the galaxies. For the Sculptor/Leo I
analogues, there are no obvious tidal features, only a very small number of
stripped star particles.
}
\label{fig:stream}
\end{figure*}

Fornax, Sculptor, and Leo I all have measured proper motions from which their
orbits can be reconstructed. These orbits are, however, subject to assumptions about the shape and amplitude of the Galactic potential. For instance, using measured radial velocities and proper motions of stars in Fornax, \citet{Piatek_etal07} estimate its
perigalacticon, $r_{p}$, and apogalacticon, $r_{a}$, adopting a Galactic potential model from \citet{Johnston_etal95}. They find $95\%$
confidence intervals of $66<r_{p}<137$~kpc and $142 < r_{a} < 242$~kpc. The
implied eccentricity of the orbit, $e  = (r_{a}-r_{p})/(r_{a}+r_{p})$,  is then 
 $0.11 < e<  0.38$. 

In Fig.~\ref{fig:dm_ms_peri} we show the pericentre distance as a function of stellar mass loss fraction (left panel) and ${V_{\rm max}}(z=0)/{V_{\rm max}}(z_{\rm infall})$ (right panel) for subhaloes with stellar mass $6\times10^6 \le M_{\star} \le 6\times10^7 \mathrm{M_{\odot}}$, which is the luminosity range of our Fornax analogues. The yellow shaded areas mark the 95$\%$ confidence interval range of orbital perigalacticon from \cite{Piatek_etal07} (with their assumed MW potential) and the grey shaded areas are the 95$\%$ confidence intervals for perigalacticon derived using the same orbital velocity measurements in combination with the potentials taken directly from the APOSTLE simulations. 

To account for the response of the Fornax motion in the APOSTLE Milky Way potentials, we derive the range of \textit{latest} orbital pericentre distance using the equations that describe motion of a bound star in a spherical potential \citep{Binney_etal08} with the observed Fornax radial motion and proper motion from \cite{Piatek_etal07}.  We assume that the potential is static. The pericentre and apocentre distance is then obtained by solving for $r$ in the following equation:
\beq
{1\over r^2} +{2(E - \phi(r)) \over L^2} =0,
\label{eq:orbit}
\eeq
where $E$ is the specific energy of the orbit, $L$ is the specific angular momentum, and $\phi(r)$ is the potential. We have verified that this procedure, assuming the spherically averaged simulated potential at $z=0$, recovers accurately the last orbital pericentre and apocentre distances of the actual orbits of the subhaloes in the same simulations. We compute the 95$\%$ confidence interval on the pericentre distance by Monte-Carlo sampling from the error distributions of Fornax distance, radial velocity and proper motion, assuming these distributions to be Gaussian.

We find that the derived orbital pericentre and apocentre ranges vary with the assumption of the underlying Galactic potential. We have considered the six HR Galactic potentials from APOSTLE and the Galactic potential model from \citet{Johnston_etal95}. We list the results of these calculations in Table 1.      

We find that a population of Fornax analogues has a pericentre distance smaller than the lower bound on that of Fornax itself, more so for the analogues experiencing the greatest stellar tidal stripping. For the analogues that do lie within the shaded regions in Fig.~\ref{fig:dm_ms_peri} (both from the MR simulations), the highest stellar mass loss fraction is $\lsim 10\%$, which corresponds to $\sim 6.7\times10^5 \,\mathrm{M_{\odot}}$ of tidally stripped stars. By interpolating along the approximate locus of the analogues in Fig.~\ref{fig:dm_ms_peri} we can infer that the observed bounds on the likely pericentre of Fornax are consistent with stellar mass loss fractions of up to $\lsim 15-20\%$. 

We note that even though a majority of the Fornax analogues are heavily tidally stripped, most of them have orbits that are not consistent with the orbital properties (e.g. pericentre distance) inferred from the measurements of the motion of Fornax. Precisely matching the real  orbital properties if Fornax is difficult, due to the small sample size in our simulations. As a result, only two subhaloes satisfy all constraints when the orbital data is included. In addition to the implications for the formation of Fornax, this may have interesting implications for the Milky Way potential. For example, in Table 1 we show that a more massive Galactic potential predicts smaller pericentre distances. If the Milky Way halo mass is less than $\sim 10^{12} \mathrm{M_{\odot}}$, the 95$\%$ confidence lower bound on the pericentre is $\sim 110$~kpc (see Table 1). This is also inconsistent with any of our Fornax analogues. 

In the case of Sculptor, a similar orbital study was carried out by
\citet{Piatek_etal06}. They obtain $95\%$ confidence intervals on
perigalacticon and apogalacticon of $31<r_{p}<83$~kpc and $97<r_{a}<313$~kpc,
implying $0.26<e<0.60$ with the Galactic potential model from \citet{Johnston_etal95}. In the case of  Leo I, a study of HST proper motion measurements was carried out
by~\citet{Sohn_etal13}. Using three different mass models for the Galactic potential with total virial masses of $1.0\times10^{12}, 1.5\times10^{12}$, and $2.0\times10^{12} \mathrm{M_{\odot}}$ respectively, they infer a perigalacticon occurring $1.05\pm0.09$~Gyr ago at a galactocentric distance of $r_{p}=91\pm36$~kpc. On
this basis they estimate that Leo I entered the Milky Way virial radius $2.33\pm 0.21$~Gyr ago and is most likely on its first infall. Most of our Sculptor/Leo I analogues lie within those perigalacticon and apogalacticon ranges, although the some of them have higher orbital eccentricities.    

\section{Tidal Debris}
\label{section:tidal_debris}

In this section we study the surface
brightness of the outer regions of the analogues, including their tidal debris,
in order to assess the detectability of tidal features. We also
investigate the possibility that tidally stripped star may contaminate
kinematic observations of real dSphs. 

\begin{figure*}
\includegraphics[height=8.1 cm]{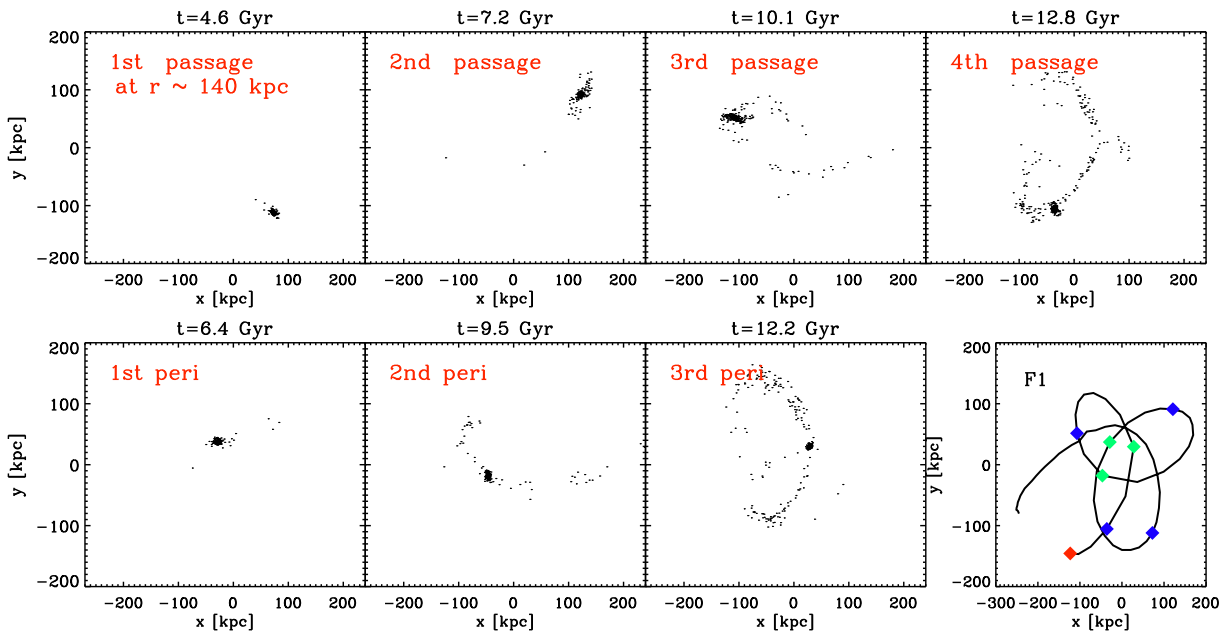}
\caption{
Projected spatial distribution of all star particles bound to F2 at infall or any time thereafter, at different times $t$. The upper panels show the star particle distribution when the galaxy is at a galactocentric distance $\sim 140$~kpc (the current distance of Fornax). The lower panels show equivalent distributions when F2 is at each of its perigalactic passages. The orbit of F2
is shown in the lower right panel. Blue points along the orbit mark the locations with a galactocentric distance of $140$~kpc and green points mark
perigalacticons. The red diamond marks the present-day location of the
satellite.
}
\label{fig:orbit}
\end{figure*}

\subsection{Do dSphs have stellar tidal tails?}
\label{subsection:tidal tials}

In Fig.~\ref{fig:stream} we show the time evolution of the stellar mass
surface density distribution for three HR dSph analogues: F1, F2, and SL2. In the
leftmost panels, which correspond to these galaxies at their times of infall,
the star particles are deeply embedded within their host haloes, as expected.
As time evolves (from left to the right), we see that tidal stripping of stars
from F1 and F2 leads to the development of leading and trailing streams that
approximately trace their obits. Our results in \S\ref{subsection:orbit} imply that Fornax may have lost up to $10-20\%$ of its infall stellar mass (equivalent to a stellar mass $\sim 7\times10^5 M_{\odot}$). Among the HR analogues\footnote{The resolution of the MR analogues is too low to study faint tidal features.} F2 is closest to having this amount of mass loss. It is therefore likely to be a much better predictor of tidal features that may be observed around Fornax than F1, which has lost $f_{*,\mathrm{loss}} \sim 58\%$ of its stars. F2 has $f_{*,\mathrm{loss}} \sim 24\%$, which is about $\sim 6.3\times10^5 \mathrm{M_{\odot}}$ in stellar mass, but this is nevertheless sufficient to generate extended coherent tidal features. In contrast, SL2 does not have clear tidal tails, with only a handful of
unbound star particles ($f_{*,\mathrm{loss}} \sim 5\%$). Indeed, as is shown in Fig.~\ref{fig:ms_loss}, most
of the Sculptor and Leo I analogues in exhibit very little stellar mass loss, usually less than $20\%$. The qualitative features of SL2 shown in
Fig.~\ref{fig:stream} are representative of such analogues.  

\begin{table*}
{\renewcommand{\arraystretch}{1.3}
\renewcommand{\tabcolsep}{0.2cm}
\begin{tabular}{l c c c c c}
\hline 
\hline
Galactic Potential &  $M_{200}$ & $r_p$ ($D=138$ kpc) & $95\%$ conf. & $r_p$ ($D=147$ kpc)& $95\%$ conf. \\
  & [$10^{12} M_{\rm \odot}$] &  [kpc] & [kpc] &  [kpc]& [kpc]\\
\hline
AP-1  G1&1.66& 128& (76, 149)& 140.0 &(93, 165)\\
AP-1  G2&1.10& 135& (113, 152)& 144.0 &(121, 167)\\
AP-4  G1&1.38& 133& (97, 151)& 144.0 &(113, 167)\\
AP-4  G2&1.35& 134& (102, 151)& 144.0 &(116, 167)\\
AP-11  G1&0.99& 135& (111, 152)& 145.0 &(121, 167)\\
AP-11  G2&0.80& 136& (114, 152)& 145.0 &(123, 167)\\
\hline
Johnston et al.(1995)&--& 125& (74, 148)& 138.0 &(90, 164)\\
\hline
\end{tabular}
\medskip
\caption{Pericentre distances ($r_\mathrm{p}$)  of Fornax derived with different assumptions regarding the Galactic potential and the current distance of Fornax. "Ap-1", "Ap-4", and "Ap-11" refer to different HR APOSTLE simulations (see \S\ref{section:simulations} for details). "G1" and "G2" indicate the two different MW-like galaxies in each simulation. Here we show pericentre predictions with two different assumptions for the current Fornax distance: $D$ = 138 or 147 kpc.}
 }
\label{tb:orbit}
\end{table*}

The upper panels of Fig.~\ref{fig:orbit} illustrate the star particle
distribution of F2 at different phases along its orbit: in the upper panel, at the observed distance of Fornax ($D = 140$~kpc), and in the lower panels at each perigalacticon. F2 has three perigalactic passages after its infall
$\sim10$~Gyr ago. Even after the first perigalacticon, the galaxy exhibits clear signs of tidal stripping.

\begin{figure*}
\includegraphics[height=8.9 cm]{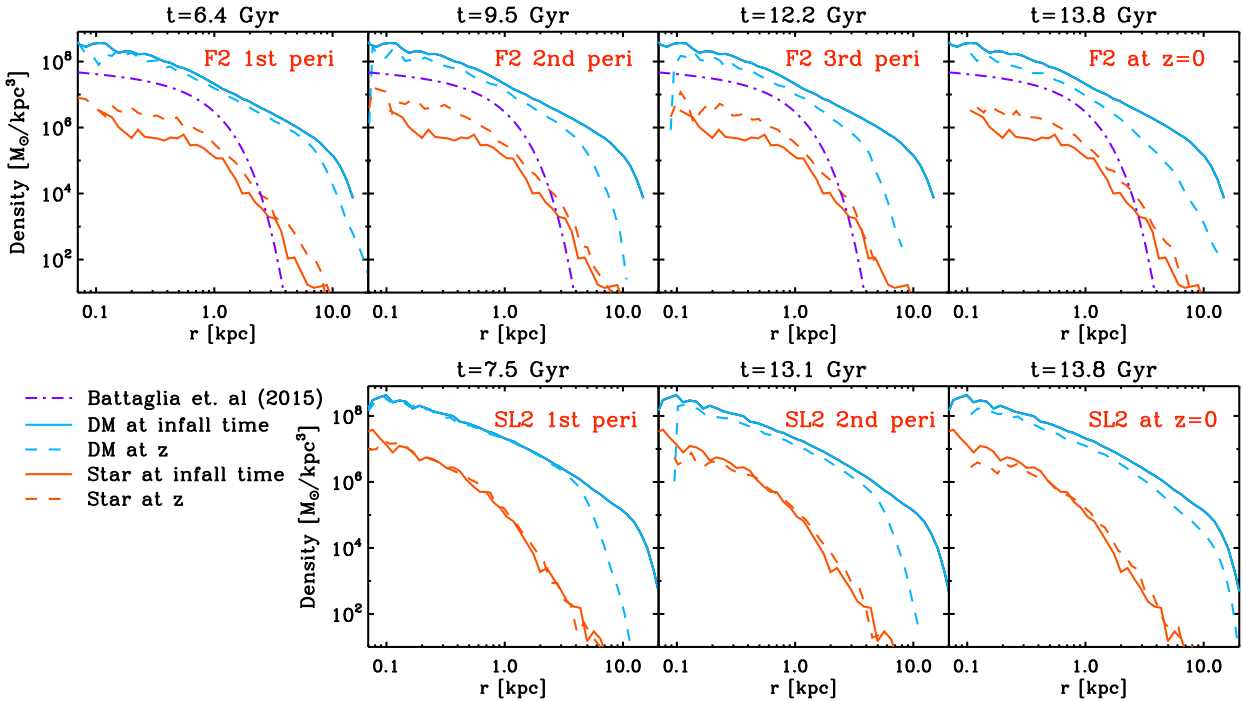}
\vspace{1em}
\caption{ 
Dark matter (blue lines) and stellar (red lines) density profiles of one Fornax analogue (F2, upper panels) and one Sculptor analogue (SL2, lower panels) at successive perigalacticons (from left to right) and at $\rm z=0$ (rightmost panels). The profiles include only the gravitationally bound particles in each system. The solid lines denote the profiles at infall (z $\sim$ 2.2 for F2 and z $\sim$ 0.86 for SL2), and the dashed lines denote the profiles at later times. For F2 we also show the initial stellar density profile from \citet{Battaglia_etal15} as a purple dash-dotted line. The dark matter profiles for both F2 and SL1 continuously become shallower and lose mass in the outer region due to tidal disruption. The stellar distribution of F2 is more extended than that of SL1. For F2 the stellar densities increase at the centre after infall, as star formation continues down to $z=0.4$.
}
\label{fig:mr}
\end{figure*}

Perhaps the most obvious reason that the majority of Fornax analogues show tidal features resulting from stellar stripping, whereas most of the Sculptor/Leo I analogues do not, is simply because the Fornax analogues must have lost more mass overall in order to be identified as such. Fig.~2 demonstrates that stellar stripping correlates with total mass loss. Similar finding is illustrated in Fig. 8 in \cite{Sawala_etal16}, where Fornax is shown to have very high stellar-to-total mass ratio and will likely be explained by tidal stripping. Another factor is the extent of the stellar distribution at the time of infall. 
In Fig.~\ref{fig:mr} we show the dark matter and stellar density profiles of F2
and SL2 at their perigalacticons and at $z=0$. At infall, their host subhaloes
have similar halo masses, $\sim4\times 10^{9} \mathrm{M_{\odot}}$. At the second perigalactic
passages ($t=9.5$~Gyr for F2, $t=13.1$ Gyr for SL2) their dark matter
distributions (for the mass that SUBFIND considers bound) are both truncated at $r\sim 10$~kpc.  Fig.~\ref{fig:stream} and Fig.~\ref{fig:orbit} clearly show that, while F2
already exhibits extended stream-like features at this second perigalacticon, SL2 has not lost significant
stellar mass. This difference arises because, at infall, the stellar distribution in F2
extends up to $r\sim 10$~kpc, whereas stars in SL2 are relatively more deeply
embedded, extending only to $r \le 7$ kpc.  As the subhaloes lose weakly-bound
dark matter to tides after infall, the subhalo tidal radius of F2 becomes comparable to the extent of its stellar distribution. The greater compactness of stars in SL2 at infall explains its apparently greater resilience to stellar stripping, despite the
overall similarity in the halo mass and total mass evolution of the two
galaxies.

\subsection{Surface densities of stellar tidal features}
\label{subsection:detection}

We now provide estimates of the (typically very low) surface
brightness of the tidal features seen in the simulations and compare these to
known tidal features. We again focus on Fornax analogue F2 at $z=0$, with a
galactocentric distance of 207~kpc, to provide an upper bound on the surface brightness estimate. 

We obtain a simple measure of surface brightness by drawing a few regions of
interest around the simulated dwarf galaxies and along their stellar tidal
tails. We use all the star particles lying within these regions, assuming a stellar mass-to-light ratio $M_{\star}/L=1$, to calculate an
average apparent surface brightness. For the purpose of this exercise we choose
a fiducial projection along radial direction between the subhalo and galactic halo centre.

\begin{figure*}
\includegraphics[height=5.2cm]{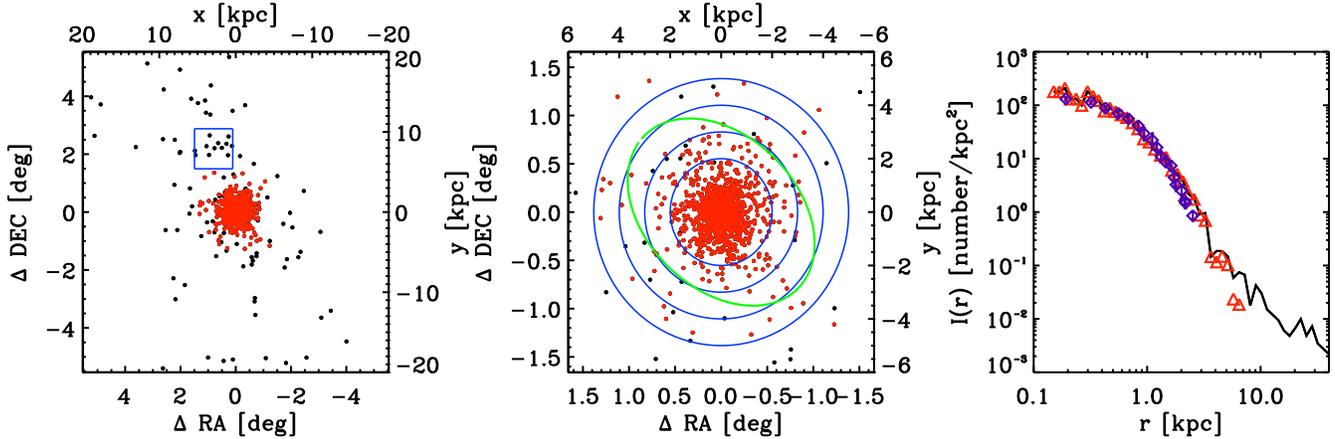}
\vspace{1em}
\caption{ 
Spatial distribution of star particles in F2 (projected along the line of sight to the centre of the main halo) at $z=0$. Red points/symbols mark the distribution of currently bound stars, and black points/lines the stripped stars. {\it Left panel:} the particle distribution within a box 40~kpc on a side ($\sim 10\deg$ for $D=204$~kpc). The average surface brightness within the blue square (5~kpc on a side) is $\sim \Sigma \sim  32.6 \sbunits$. {\it Middle panel:} a zoom-in view of the galaxy F2 (12~kpc on a side). We draw annuli (blue circles) around the galaxy with radii 2--5~kpc ($\sim0.8$--$2.0\deg$) and 1~kpc spacing to estimate surface brightness (see text) around the stellar 'tidal radius' of Fornax (i.e.\ the apparent extent of the stellar distribution, $r_{t} = 69.1\arcmin$ \citep{Battaglia_etal06}, shown as a black dashed ellipse).  {\it Right panel:} projected stellar number density profile for F2. The black line shows the total density of bound \textit{and} stripped stars together. The red triangles mark the density for bound stars only. The purple diamonds (with 1$\sigma$ error) show  the observed stellar surface density of Fornax  from \citet{Coleman_etal08}.
}
\label{fig:ir}
\end{figure*}

In Fig.~\ref{fig:ir} we show the spatial distributions of
gravitationally bound star particles in F2 (red points) and star particles that were previously
bound to F2 but have been unbound by tides (black points) at $z=0$. In the left panel we
show all the star particles lying within a $40\times40$~kpc box projected
along the $z$ axis of the simulation coordinate system. We can clearly see the tidally stripped stars extending
along the $y$ direction. Within a $5\times5$~kpc box about 2~degrees away from
the galaxy, the surface brightness of an overdense region is $\Sigma \sim 32.6 \,\sbunits$. The enclosed stellar mass within this region is $
M_{\star} \sim 7.0\times10^4 \, \mathrm{M_{\odot}}$. For comparison, the surface brightness of the
Virgo Overdensity discovered by SDSS is $\ \Sigma_r \sim 32.5 \,\sbunits$ \citep{Juric_etal08}, and the surface brightness of the
Sagittarius dwarf Northern stream is  $
\Sigma_{\mathrm{V}} \sim 31 \,\sbunits$ \citep{Martnez-Delgado_etal04}. The stellar
overdensity in Eridanus-Phoenix recently discovered by DES has $\Sigma_\mathrm{r} \sim 32.8 \,\sbunits$ \citep{Li_etal16}. These magnitudes are comparable to the surface brightness of the tidal feature from F2, therefore similar features could be detected in ongoing or existing imaging surveys. 

We also examine the stellar structure at the edge of the F2 galaxy. As shown in the middle panel of Fig.~\ref{fig:ir}, we draw 4 annuli centred on the galaxy with radius of  2--5~kpc ($\sim 0.55\deg$ -- $1.4\deg$) and 1~kpc spacing. Using naive visual inspection, there is no obvious stellar tidal structure or elongation of the body of the galaxy within 5~kpc, although this may simply be because the resolution of the simulation is not sufficient to capture any faint structures. The average surface brightness of {the annulii are}  $\Sigma \sim 31.23 \,\sbunits$ (2-3 kpc), $\Sigma \sim$ 32.8 mag $ {\rm arcsec^{-2}}$ (3-4 kpc), and $\Sigma \sim 33.6 \,\sbunits$ (4-5 kpc). The surface brightness rapidly drops below the reach of current surveys beyond $r > 3$--$4$~kpc ($r > 50.5\arcmin$--$67.4\arcmin$ for $D = 204$~kpc, or $r >73.7\arcmin$--$98.2\arcmin$ for $D=140$~kpc). This is consistent with the stellar `tidal radius' of a King-model fit to the light profile of Fornax, $r_{t} = 69.1\arcmin$ \citep{Battaglia_etal06}, beyond which the stellar distribution is indistinguishable from the background. 

We have also demonstrated in the right panel of Fig.~\ref{fig:ir} that the shape of F2's light profile agrees well with observations of Fornax itself \citep[blue diamond points;][]{Coleman_etal05}. Although there are stars gravitationally bound to F2 at $r > 3$--$4 $~kpc, their density would be indistinguishable from the background, at least with current imaging surveys. We show the surface number density of bound (red) and unbound (black) star particles in the right panel of Fig.~\ref{fig:ir}. The projected stellar density profile in the inner region of F2 is well-resolved and its shape agrees well with that of Fornax overall (we have rescaled the amplitude of the simulated profile to match approximately that of the observed data, which is based on counts of giant stars). There is a steeper drop at $r > 1.2$~kpc in the observed profile. The bound star particles in F2 extend up to $r \sim 9$~kpc and continue as a diffuse tidally stripped stellar tail.  

We note that stellar tidal features are usually identified in surveys by `matched filtering', using colour and magnitude cuts based on theoretical or empirical model isochrones to isolate, for example, main-sequence turn-off (MSTO) or red giant branch (RGB) stars from the background \citep{Rockosi_etal02,Belokurov_etal06}. Here we do not implement such a procedure, assuming instead that all stars in the tidal features are detectable. Our numbers therefore serve as upper limits on surface brightness. Due to the limited particle resolution, we can only report {\it average} surface brightness estimations over large areas, because the fine structures of tidal features are not resolved in our simulations. In reality, if there are any overdense structures around MW dSphs, they could contain smaller-scale fluctuations brighter than our estimates. We nevertheless find that there are several features along our simulated dSph tidal streams that have surface densities comparable to known stellar overdensities. The presence of such structures around galaxies such as Fornax can, therefore, be constrained in the near future by wide-field imagining observations.

\subsection{Stellar kinematic contaminations from tidal debris}
\label{subsection:contamination}

In systems with very high mass-to-light ratios, such as the MW dSphs, stellar kinematics provide a direct probe of the distribution of dark matter on very small scales. Such measurements are now one of the principle tests of the validity of the CDM model. However, the samples of stars used to study the kinematics of MW dSphs can be contaminated by tidally stripped stars that are not in equilibrium with the system. This contamination will affect the precision of dynamical mass estimates based on stellar velocities \citep{Klimentowski_etal07}. Here we demonstrate that gravitationally unbound stars are indeed likely to be confused with bound member stars in conventional stellar kinematic observations. We also demonstrate that, depending on the projected orientation of tidal tails with respect to the line-of-sight,  these unbound stars will bias stellar velocity dispersion measurements. 

\begin{figure*}
\includegraphics[height=6.7cm]{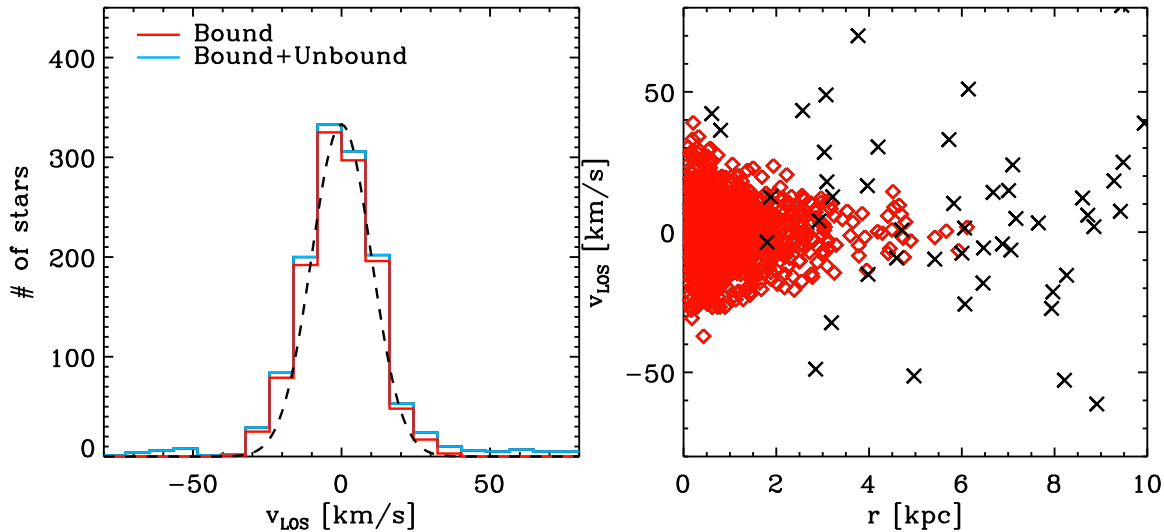}
\vspace{1em}
\caption{ 
Line-of-sight (LOS) velocity distribution of stars (velocity relative to the subhalo motion) for the F2 subhalo. The LOS direction is along the radial vector from the galactic centre to F2. {\it Left Panel}: Histograms of stars that are currently gravitationally bound to F2 (red) and those that have ever been bound to F2, at infall or any time thereafter (blue). A Gaussian distribution with width $10 \, \mathrm{km\,s^{-1}}$ is shown as a black dashed line. {\it Right Panel}: LOS velocity distribution as a function of radius. Red diamonds are bound stars, and black crosses are stripped stars.
}
\label{fig:vel_dist}
\end{figure*}

In the left panels of Fig.~\ref{fig:vel_dist} we show the line-of-sight velocity distribution of stars (red: bound only, blue: both bound and unbound) in the F2 subhalo, projected along the radial direction (i.e.\ along the vector from the centre of the galactic halo to the centre of F2) with velocity measured relative to the peculiar velocity of F2. The line-of sight velocity distributions are close to Gaussian, centred at zero relative velocity with average dispersion $\sim 10  \,   \kms$ (black dashed curves). In the right panel we plot the velocity distribution as a function of radius for both the bound (red diamonds) and stripped stars (black crosses). Some of the stripped stars have velocities much larger than those of bound stars at the same radius. However, a good fraction of these lie within the range of the velocity distribution of bound stars, between $-40$ and $40\, \kms$. 

Fig.~\ref{fig:vel_dist} indicates that a simple, non-iterative, cut-off in velocity, as commonly applied to kinematic samples  \citep[e.g.][]{Walker_etal06}, may be subject to kinematic contamination from tidally stripped stars. There are other methods that can be combined to determine membership probabilities for stars which utilize information such as the spatial distribution, the colour-magnitude diagram (CMD), the metallicity of stars, and the full velocity distribution \citep[e.g.][]{Simon_etal07,Walker_etal09b}. However, since stripped stars were associated with the target galaxies in the past, their metallicities and CMDs are likely to be similar to those of the stars that remain bound. The spatial distribution may be more informative, because stars closer to the centres of galaxies are more likely to be bound, although this method can only provide relative probabilities. Stripped stars in the foreground or background may still contaminate the velocity sample at small projected radii (see right panel of Fig.~\ref{fig:vel_dist}).  

\begin{figure*}
\includegraphics[height=7.1cm]{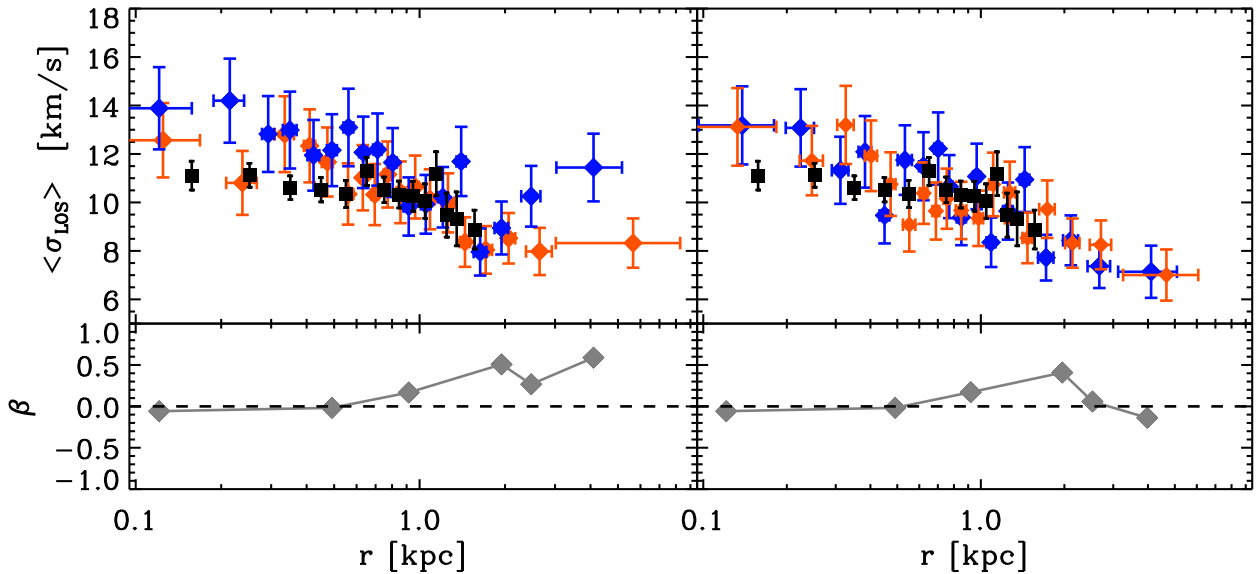}
\vspace{1em} 
\caption{ 
Line-of-sight (LOS) velocity dispersion (upper panels) and velocity anisotropy (lower panels) of stars in Fornax analogue F2. All the stars are selected to have LOS velocities in the range $\pm 40\, \mathrm{km\,s^{-1}}$ relative to the LOS velocity of F2 as a whole. The right-hand panels show results for stars that are currently gravitationally bound to F2, while the left-hand panels show results of all stars that have ever been bound to F2 at any time since its infall. We plot the LOS velocity projected along the radius vector from the galactic centre to F2 (blue points) and, for comparison, along a direction perpendicular to that vector (orange points).
}
\label{fig:vd}
\end{figure*}

In Fig.~\ref{fig:vd} we show that the unbound stars have characteristic velocity signatures. They can inflate the stellar velocity dispersion and change the stellar velocity anisotropy, to an extent that depends on the orientation of the tidal stream and the peculiar velocity of the galaxy along the line-of-sight direction. Similar results have been obtained by several previous studies \citep[e.g.\ ][]{Read_etal06,Klimentowski_etal07}. For instance, ~\cite{Klimentowski_etal07} studied $N$-body simulations of a dwarf galaxy on a highly eccentric orbit. They found that  when the dwarf is observed along its tidal tails, kinematic samples are most strongly contaminated by unbound stars from the tails (e.g.\ Figs. 5 and 6 in their paper). In the upper panels of Fig.~\ref{fig:vd}, we plot the velocity dispersion for F2 along the line-of-sight (blue points) and in a direction perpendicular to the line-of-sight (orange points). We only include stars with LOS velocities $\pm 40 \,\mathrm{km\,s^{-1}}$ relative to the peculiar velocity of F2, as these would be identified as possible member stars using the width of the velocity distribution alone. We find the velocity dispersion is inflated by unbound stars at $r > 3$~kpc by a factor of $>30\%$ (depending on the LOS projection) relative to the velocity dispersion of bound stars at $r \sim 3$~kpc. 

In the lower panels of Fig.~\ref{fig:vd} we show the velocity dispersion anisotropy $\beta(r)$, which denotes the ratio between the tangential and radial components of the velocity dispersion. At a given radius,
\beq
\beta(r)=1-{\sigma_t^2 \over 2\sigma_r^2} ,
\label{eq:ansi}
\eeq
where $\sigma_t^2$ is the tangential component of the velocity dispersion, and $\sigma_r^2$ is the radial component. In F2, the velocity anisotropy for the bound stars (lower right panel) is consistent with zero at most of the radii, with a variation between $-0.1$ and $0.5$. This indicates that isotropic orbits are good approximations for bound stars in the F2 galaxy. However, at $r > 2$~kpc, where unbound stars start to dominate, the velocity anisotropy can deviate significantly from zero. For example, the curve in lower left panel shows a increase in anisotropy. This result can be explained by the large radial motions of stripped stars along the stellar stream.


\subsection{Comparison to previous work on Fornax tidal evolution}
\label{subsection: comparison}

\citet{Battaglia_etal15} construct a set of N-body simulations to study the effect of tides on Fornax. However, contrary to our results, they do not find any evidence for stellar tidal features. To understand this discrepancy, we compare the initial conditions of their simulations with our Fornax analogues. They adopted two orbital setups: one with perigalactic radius $r_p=118$~kpc and apogalactic radius $r_a = 152$~kpc, yielding an orbital eccentricity $e = 0.13$, and another with $e = 0.4$ and $r_p = 61 $~kpc. Since the orbit of our subhalo F2 is quite similar to their second orbital configuration (first $r_p \sim 68$~kpc and $r_{a} \sim 153$~kpc after infall), we can conclude that the difference is not entirely due to the  orbits. 

The initial DM mass adopted in~\citet{Battaglia_etal15} was $3\times 10^9\, \mathrm{M_{\odot}}$, which is similar to the infall mass of F2, $4 \times 10^9 \mathrm{M_{\odot}}$. The infall stellar mass in our simulation is lower than their initial stellar mass ($5\times 10^7 \mathrm{M_{\odot}}$ versus our value of 3$\times 10^6 \mathrm{M_{\odot}}$). In addition, in order to increase the possibility of generating stellar tidal disruption, \citet{Battaglia_etal15}  assumed a cored DM density profile. 

Therefore purely on the basis of its subhalo mass and DM density profile, our Fornax DM subhalo should be more resilient to tidal forces than their subhalo. However, when we compare the stellar density profiles, which are shown in Fig.~\ref{fig:mr}, we see that their stellar distribution is much more concentrated than ours. In the upper panels of Fig.~\ref{fig:mr}, the purple dash-dotted lines show the stellar density profile of \citet{Battaglia_etal15}. It declines very rapidly around Fornax's nominal King stellar `tidal radius' of $\sim 3$ kpc. In contrast, our profiles have an extended distribution up to $\sim 7$~kpc (at $z=0$). However, as shown in \S~\ref{subsection:detection}, the surface brightness of F2 drops quickly at $r> 3$~kpc, at which point it falls below the limiting background density of current observations. It is therefore quite likely that, in reality, the Fornax stellar distribution extends far beyond the apparent `tidal radius'.  This may explain why our conclusions regarding tidal features around Fornax differ from those of \citet{Battaglia_etal15}.
 
\section{Conclusion and Discussion}
\label{section:conclusion}

We have used the high-resolution APOSTLE hydrodynamical simulations to investigate, for the first time, the stellar tidal stripping of classical MW dSphs in the range of $M_{\star} = 10^6-10^8 \, \mathrm{M_{\odot}}$ in a cosmological context. This cosmological context is important because the factors that determine the severity of tidal effects, including the assembly history and structure of the potential well of the host galaxy, the mass and structure of the satellite galaxies, and their infall times, stellar content and orbits, are all self-consistently taken into account. In these simulations, we find many examples of stellar stripping of satellites due to tidal interactions with their host; $\sim$ 30$\%$ of satellites lose more than 20$\%$ of their stellar mass between their time of infall and the present day. This fraction is remarkably high given that only a handful of MW dSphs display clear signs of tidal tails or strong structural distortions that could be attributed to tidal stirring. However, given the difficulty of detecting diffuse low-surface brightness tidal features, the apparent contradictions can be explained by a lack of sufficiently deep wide-area photometric data. Ongoing and future deep imaging surveys, such as DES and LSST, will be able to verify this in the near future.

We have identified Fornax analogues that best match the observed orbital properties of Fornax. We note that many of our Fornax analogue subhaloes are severely tidally stripped, but they tend to have lower orbital pericentre distance than the predictions derived from measurements of the velocity of Fornax. This puzzling tension may indicate that  Fornax's orbit is in some way unusual, or provide constraints on the Galactic potential. We leave this work to a future study with more detailed modeling of the Galactic potential and orbital simulations of Fornax.

Nevertheless, those Fornax analogues that match the orbital properties have typically lost $\sim10-20\%$ of their infall stellar mass and exhibit faint stellar tidal streams along their orbit. This tidal debris is generally low in surface brightness, but with overdense regions that are comparable to currently known stellar overdensities and streams in the MW.  We also show that these Fornax analogues may have diffuse but nevertheless bound stellar components extending beyond the conventional Fornax stellar `tidal radius'. These extended envelopes are difficult to distinguish from background stars. 

Our results are interesting in the context of recent observational and theoretical studies of Fornax. For example~\cite{Bate_etal15} have used VST ATLAS survey data to show that a population of Fornax RGB stars extends all the way out to the Fornax stellar `tidal radius' defined by \citet{Battaglia_etal06}. This observation is in line with the predictions of our simulation, and differs from the conclusions of~\citet{Battaglia_etal15} that stars Fornax do not extended beyond that radius, and that Fornax has undergone no significant stellar tidal stripping. The lack of significant stellar tidal features in our Sculptor and Leo I analogues can, however, be explained by their compact stellar distribution, even though their dark matter tidal radius (different than the stellar one) may be similar to that of the Fornax analogues after a few pericentric passages.

Having identified Fornax, Sculptor and Leo I analogues in the simulations on the basis of their observed stellar kinematics and luminosities, we find that they exhibit different mass assembly histories and star formation histories. Fornax analogues in particular suffer significant mass loss, with early infall $\sim 8$-$10$~Gyr ago. They exhibit continuous star formation activity after infall, although this activity has generally ceased by the present day. On the other hand, the mass loss of Sculptor and Leo I analogues is much less severe. These analogues tend to have more recent infall times than Fornax, with star formation suppressed very rapidly after infall.

As noted by~\cite{Wang_etal16}, the correlation between infall time and star formation timescale could also be useful in testing the nature of DM, because different DM models predict different infall times. As an example, in the case of Fornax, CDM models always predict infall at lookback times of $\sim 8$-$9$~Gyr ago, whereas alternatives, such as warm DM (WDM) or decaying DM (DDM) models, predict infall times as recent as $3$-$4$ Gyr ago. It is interesting to note that Fornax does show signs of recent star formation activity \citep{Weisz_etal15}, $\sim 2$ Gyr ago. This short timescale for shutting down star formation may be in tension with infall $> 5$~Gyr ago. Thus it is important to understand how the star formation timescales of dSphs can be used to constrain their likely orbital history, which in turn may enable stronger tests of the nature of DM. 

Finally, we have discussed the potential for kinematic contamination by tidally stripped stars in the context of dSph stellar velocity dispersion measurements. We have shown that unbound stars can inflate the stellar velocity dispersion and change the average stellar velocity anisotropy in the outskirt of dSphs. We find that a conventional cut on heliocentric velocities cannot exclude all the unbound particles, nor can metallicity cuts or selections based on CMD filtering, since the contaminating stars were once gravitationally-bound member stars. On the other hand, these kinematic features can also serve as a signature to detect tidal debris.

%
%
\section*{Acknowledgments}
We would like to thank Nitya Kallivayalil, Manoj Kaplinghat, Mike Boylan-Kolchin, Robyn Sanderson, Josh Simon, and Ting Li for useful discussions. MYW and LES acknowledge support from NSF grant PHY-1522717. APC is supported by a COFUND/Durham Junior Research Fellowship under EU grant 267209 and by STFC (ST/L00075X/1). T. S. acknowledges the support of the Academy of Finland grant 1274931. JFN was supported by a Visiting Professorship Grant from the Leverhulme Trust VP1-2014-021. This work used the DiRAC Data Centric system at Durham University, operated by the Institute for Computational Cosmology on behalf of the STFC DiRAC HPC Facility (www.dirac.ac.uk). This equipment was funded by BIS National E-infrastructure capital grant ST/K00042X/1, STFC capital grants ST/H008519/1 and ST/K00087X/1, STFC DiRAC Operations grant ST/K003267/1 and Durham University. DiRAC is part of the National E-Infrastructure. 

\bibliography{dSph_tidal}


\end{document}